\documentclass[twocolumn,showpacs,amsmath,amssymb,epsfig]{revtex4}
\usepackage{graphicx}
\usepackage{bm}
\usepackage{epsfig}
\makeatletter
\newcommand\figcaption{\def\@captype{figure}\caption}
\newcommand\tabcaption{\def\@captype{table}\caption}
\makeatother
\begin{document}
\title{Scaling of Nonlinear Longitudinal and Hall Resistivities near the Vortex Glass Transition}
 \draft
\author{ X. Hu, L. He, L. Yin, Z.H. Ning, H.Y. Xu, X.L. Xu, J.D. Guo, C.Y. Li and D.L. Yin*}
\address{Department of Physics, Peking University, Beijing 100871, China}
\date{\today}
\begin{abstract}
We show that the longitudinal current-voltage characteristics of
superconductors in mixed state have the general form of extended
power law. Isotherms simulated from this nonlinear equation fit
the experimental I-V data of Strachan et al. [ Phys. Rev. Lett.
{\bf 87}, 067007 (2001)]. We determine the average pinning force
in the flux creep and strong pinning regime and discuss both the
puzzling scaling behavior $\rho_{xy}\propto\rho_{xx}^\beta$ and a
recently found new scaling relationship of nonlinear Hall
resistivity $\rho_{xy}(T)$.
\end{abstract}
\pacs{74.25.Sv,74.25.Fy, 74.25.Ha,74.20.De} \maketitle

The vortex motion in the superfluid electrons has presented a
persistent problem in the physics of high-temperature
superconductors(HTS). One of the most puzzling phenomena is the
sign change that has been observed in the Hall effect in the
superconducting state in most HTS and some conventional
superconductors \cite{Hagen}. Another highly intriguing issue is
the power law scaling relation between the Hall resistivity
$\rho_{xy}$ and longitudinal resistivity $\rho_{xx}$
\cite{Luo,Kang}. Experiments show that the Hall effect at least
depends on two factors: the doping level \cite{Nagaoka} and the
vortex pinning \cite{Luo,Woltigens, Kopnin}. Based on the
hypothesis of a vortex glass transition \cite{Fisher}, Dorsey and
Fisher \cite{Dorsey} reasonably explained the puzzling scaling
behavior $\rho_{xy}\propto[\rho_{xx}(T)]^{\beta}$ with
$\beta=1.7\pm0.2$ observed by Luo et al. \cite{Luo}. The Hall
effect itself was attributed to a "particle-hole" asymmetry and
the exponent $\beta$ was related to a specially chosen
particle-hole asymmetry exponent $\lambda\simeq3$. They predicted
"The nonlinear Hall electric field $E_y$ should exhibit universal
scaling and, right at the transition, should vanish with a
universal power of the current density $J_x$". This prediction has
been verified by W\"{o}ltigens et al. \cite{Woltigens} with an
experimentally determined Hall-related exponent
$\lambda=3.4\pm0.3$. Furthermore, they found wide range scaling
behavior $\rho_{xy}=A[\rho_{xx}(J_x,T)]^{2.0\pm0.2}$ in
consistency with the model proposed by Vinokur, Geshkenbein,
Feigel'man and Blatter(VGFB) \cite{Vinokur}. However, recently the
widely accepted experimental evidence for vortex glass theory
\cite{Koch} has met serious doubt \cite{Strachan}. Strachan et al.
show wide range accurate isothermal I-V measurements over 5 or 6
decades on a high quality 2200 $\text{\AA}$ thick
YBa$_2$Cu$_3$O$_{7-\delta}$ film with $T_{c}\approx 91.5K$ and
transition width about 0.5K \cite{Strachan} and find although the
I-V isotherms measured in a magnetic field can be collapsed into
scaling functions proposed by Fisher et al. \cite{Fisher} as is
widely reported in the literature, these excellent data collapse
can also be achieved for a wide range of exponents and glass
temperatures $T_{g}$ as demonstrated in their Fig.2(a),2(b) and
2(c) \cite{Strachan}. Since the critical temperature $T_g$ cannot
be determined uniquely, the correctness of the vortex glass
picture has to be reinvestigated. In this work, we show that the
longitudinal isothermal current-voltage characteristics have the
general form of extended power law \cite{Ning}. This remarkable
feature naturally explains the insightful new findings of Strachan
et al. \cite{Strachan} as well as the recently observed scaling
relationship of $\rho_{xy}(T)$.

Thermally activated flux motion can be considered as the sequence
of thermally activated jumps of the vortex segments or vortex
bundles between the metastable states generated by disorder. Every
elementary jump is viewed as the nucleation of a vortex loop, and
the mean velocity of the vortex system is determined by the
nucleation rate \cite{Fisher, NelsonVinokur}
\begin{equation}\label{eq2}
v\varpropto \exp \left(- \delta F/kT\right)\text{.}
\end{equation}

Here $\delta F$ is the free energy for the formation of the
critical size loop or nucleus which can be found by means of the
standard variational procedure from the free energy functional due
to the in-plane displacement ${\bf u}\left(z\right)$ of the moving
vortex during loop formation
\begin{equation}\label{eq3}
F_{ loop}\left[{\bf u}\right] =\int dz\left[ \frac 12%
\varepsilon _l\left| \frac{d{\bf u}\left( z\right) }{%
dz}\right| ^2+V_P\left({\bf u}\left(z\right)\right)-{\bf
f_s}\cdot{\bf u}\right]
\end{equation}
with
\begin{equation}\label{eq4}
{\bf f_s}={\bf f_L}+{\bf f_\eta}=\frac{\Phi _0}{c}{\bf J}_p \times
{\bf e_z}
\text{\ \ and\ \ }
J_p=J-J_f=J-\frac E{\rho _f}\text{,}
\end{equation}
where ${\bf f_L}=\Phi_0{\bf J}\times{\bf e_z}/c$ is the Lorentz
force due to applied current $J$ and $f_\eta $ is the viscous drag
force on vortex, ${\bf f_\eta}=-\eta{\bf v_{vortex}}$, with ${\bf
v_{vortex}}=d{\bf u}/dt$ and viscous drag coefficient $\eta
\approx \left( \Phi _0B_{c2}\right)/\left( \rho _nc^2\right)$,
$\rho_{f}$ is the flux flow resistance of a pinning free mixed
state $\frac{\rho _f}{\rho _n}\approx \frac B{B_{c2}}\text{,}$ as
derived by Bardeen and Stephen \cite{Bardeen}.

Our equation (\ref{eq3}) differs from the corresponding one used
in previous vortex glass theories on that we considered both the
Lorentz force ${\bf f_L}$ and the viscous drag force ${\bf
f_\eta}$ on the vortex loop during its formation while in present
vortex glass theories only bare Lorentz force ${\bf f_L}$ has been
taken into account \cite{Fisher, Blatter, NelsonVinokur}.

As shown in \cite{Tinkham}, the barrier energy found by the
conventional vortex glass model \cite{Fisher} has the general form
with that of collective pinning model and Boson glass model

\begin{equation}\label{eq27}
\delta F\equiv F_{loop}[L^*(J)]=U(J)\approx U_0(\frac{J_0}{J})^\mu
\text{,}
\end{equation}
with $L^*(J)$ the critical size of vortex loop nucleation. Since
the viscous drag force ${\bf f_{\eta}}$ has been taken into
account in Eq.(\ref{eq3}), now instead of Eq.(\ref{eq27}) one must
have the barrier energy
\begin{equation}\label{eq6}
\delta F=U(J_p)\approx U_0(\frac{J_0}{J_p})^\mu \text{,}
\end{equation}
which implies a current-voltage characteristic of the form
\begin{equation}\label{eq11}
E(J)=\rho_{f}J\exp{[-\frac{U_{0}}{kT}(\frac{J_{0}}{J_{p}})^{\mu}]}
\text{,}
\end{equation}
where $U_{0}$ is a temperature and field dependent characteristic
pinning energy related to the stiffness coefficient and $J_{0}$ is
a characteristic current density related to $U_0$, $\mu$ is an
numerical exponent.

Considering the real sample size effect we note that the barrier
energy for vortex loop excitation cannot exceed the free energy
value needed for generating a loop with real sample size D defined
as
\begin{equation}\label{eq7}
\delta F=U(J)\leq U(J_D)\equiv F_{loop}[L^*(J=J_D)\equiv
D]\text{.}
\end{equation}

Therefore, the free energy barrier for vortex loop excitation
never diverges even at vanishing applied current $J$ as predicted
by the conventional vortex glass theory \cite{Fisher}.

Replacing $J_p$ in Eq.(\ref{eq11}) with the expression
Eq.(\ref{eq4}), and taking the logarithm of its both sides, one
obtains
\begin{eqnarray}\label{eq12}\nonumber
J-J_f&=&\left(\frac{U_0}{kT}\right)^{1/\mu}J_0\left[\ln\left(\frac
J{J_f}\right)\right]^{-1/\mu}\\&=&\left(\frac{U_0}{kT}\right)^{1/\mu}J_0(1+h)^{-1}\left[\ln
\left(\frac{J_D}{J_{Df}}\right)\right]^{-1/\mu}\hspace{-4mm}\text{,}
\end{eqnarray}
where $J_{Df}\equiv E(J_D)/\rho_f$ is much smaller than $J_D$ and
we define
$h\equiv\frac{-[\ln(J_D/J_{Df})^{1/\mu}-\ln(J/J_f)^{1/\mu}]}{\ln
(J_D/J_{Df})^{1/\mu}}$. Using the approximation
$(1+h)^{-1}\approx1-h$ for $|h|<1$, one finally obtains a general
normalized form of the current-voltage characteristic as
\begin{equation}\label{eq16}
y=x\exp \left[ -\gamma \left( 1+y-x\right) ^p\right]\text{,}
\end{equation}
where
\begin{equation}
\gamma =\frac{U_0}{kT}\left(\frac{2J_0}{J_D}\right)^{\mu} \text{,}
\begin{array}{lll}
&&
\end{array}
\hspace{-2mm}x=\frac J{2J_D}\text{,}
\begin{array}{lll}
&&
\end{array}
\hspace{-2mm}y=\frac{E\left( J\right) }{2\rho _fJ_D}\text{,}
\begin{array}{lll}
&&
\end{array}
\hspace{-2mm}p=\mu\\\nonumber\text{.}
\end{equation}

Equation(\ref{eq16}) can also be nearly equivalently expressed in
an extended power law form
\begin{equation}\label{eq17}
\frac{y}{x}= e^{-\gamma}(x-y+1)^{n-1}\text{,}
\end{equation}
with n is a parameter determined by the condition that
Eqs.(\ref{eq16}) and (\ref{eq17}) meet at the inflection points of
$lny\sim lnx$ curves for Eq.(\ref{eq16}).

In Fig.1(a) we show the numerical solutions of Eq.(\ref{eq16}) and
Eq.(\ref{eq17}) for comparison.

The complex longitudinal experimental current-voltage data are
usually described phenomenologically with a form \cite{Blatter}
\begin{equation}\label{eq26}
E(J)=J\rho_{xx}(T,B,J)=J\rho_{f}e^{-U_{eff}(T,B,J)/kT}\text{.}
\end{equation}
Equation(\ref{eq16}) implies that effective barrier can be
explicitly expressed as
\begin{equation}\label{eq19}
U_{eff}(T,B,J)=U_{c}(B,T)F[J/J_{c}(B,T)]\text{,}
\end{equation}
with $J_{c}(B,T)=2J_{D}(B,T)$.

Incorporating it into the commonly observed scaling behavior of
magnetic hysteresis $M(H)$ in superconductors, it can be shown
that $U_{c}(B,T)$ and $J_{c}(B,T)$ in Eq.(\ref{eq19}) must take
the following forms \cite{Blatter}
\begin{eqnarray}\label{eq20}
U_{c}(B,T)=\Psi(T)B^{n}\propto[T^{*}(B)-T]^{\delta}B^{n}\text{,}\nonumber\\
J_{c}(B,T)=\lambda(T)B^{m}\propto[T^{*}(B)-T]^{\alpha}B^{m}\text{,}
\end{eqnarray}
with $T^{*}(B)$ being the irreversibility temperature where tilt
modules and stiffness of vortex matter vanish. Considering
$\rho_{f}\propto T$, one finds from Eq.(\ref{eq19}) the following
temperature dependent relations for $\gamma$ and $I-V$ isotherms
of optimally doped YBCO samples in a given magnetic field as
\begin{eqnarray}\label{eq21}
& &\gamma(T)=\gamma_{0}(T^{*}-T)^{\delta}/kT\text{,}\nonumber\\
& &I\propto xJ_{D}(T)\propto x(T^{*}-T)^{\alpha}\text{,}\nonumber\\
& &V\propto yJ_{D}(T)\rho_{f}\propto y(T^{*}-T)^{\alpha}T\text{.}
\end{eqnarray}

The combination of equations (\ref{eq16}) and (\ref{eq21}) enables
us to reproduce the experimentally measured $I-V$ isotherms of
Strachan et al. \cite{Strachan}. In Fig.1(b) we show the
comparison, where the parameters used are $\alpha=3.0, \delta=0.5,
p=1.75, \gamma_{0}=3.21\times10^{-21}$ and the irreversibility
temperature $T^{*}(B=4T)$ is assumed $88K$ similar to the Fig.1 in
Ref.\cite{Strachan}.
\begin{figure}[htb]

\centering

\leftskip4mm \vspace{-3mm}

\includegraphics[width=0.80\linewidth]{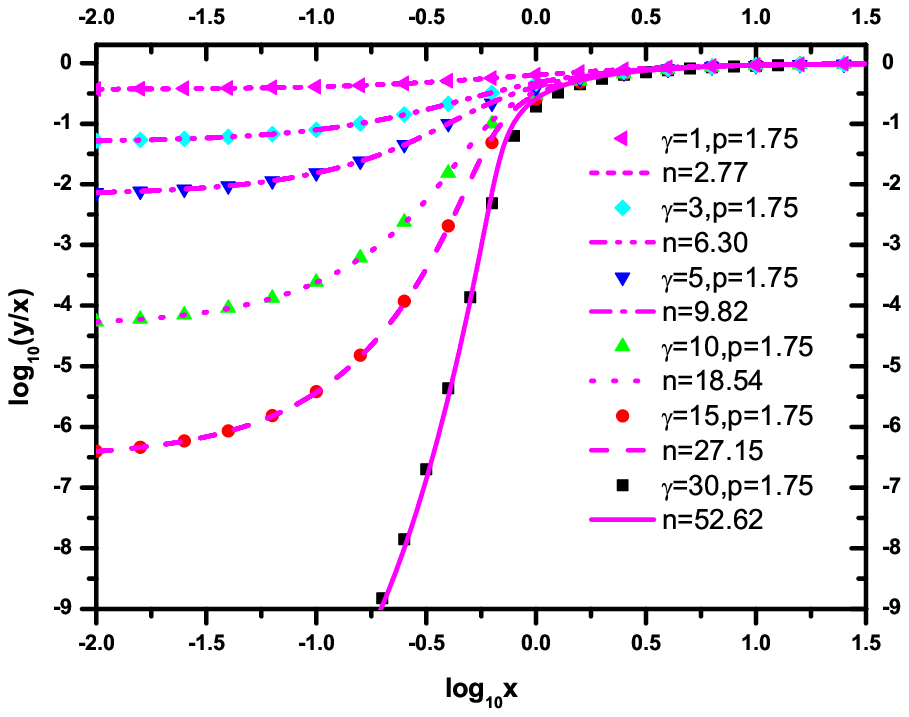}

\vspace{-3mm}

\hspace{33mm}(a)

\includegraphics[width=0.80\linewidth]{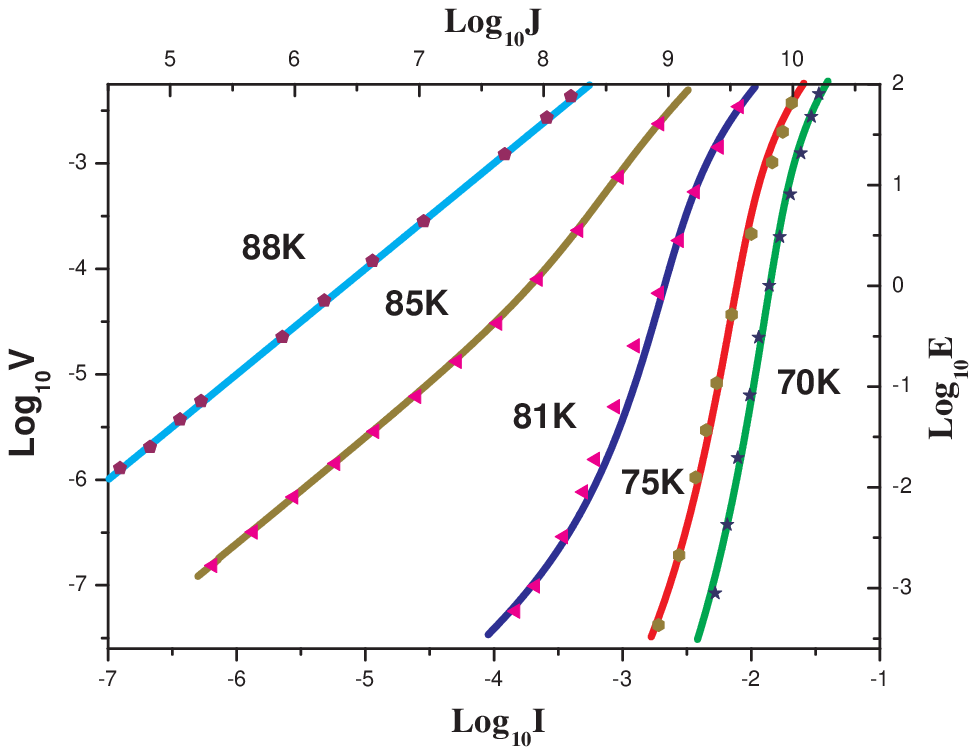}

\vspace{-3mm}

\hspace{33mm}(b)

\caption{(a)Comparison of solutions of Eq.(\ref{eq16})(solid
symbols) and Eq.(\ref{eq17})(curves).\\
(b)Comparison of our simulated I-V isotherms(lines) with the
experimental data of Strachan {\it et al.} \cite{Strachan} (solid
symbols).}
\end{figure}

\begin{figure}[htb]

\centering

\includegraphics[width=0.585\linewidth,angle=270]{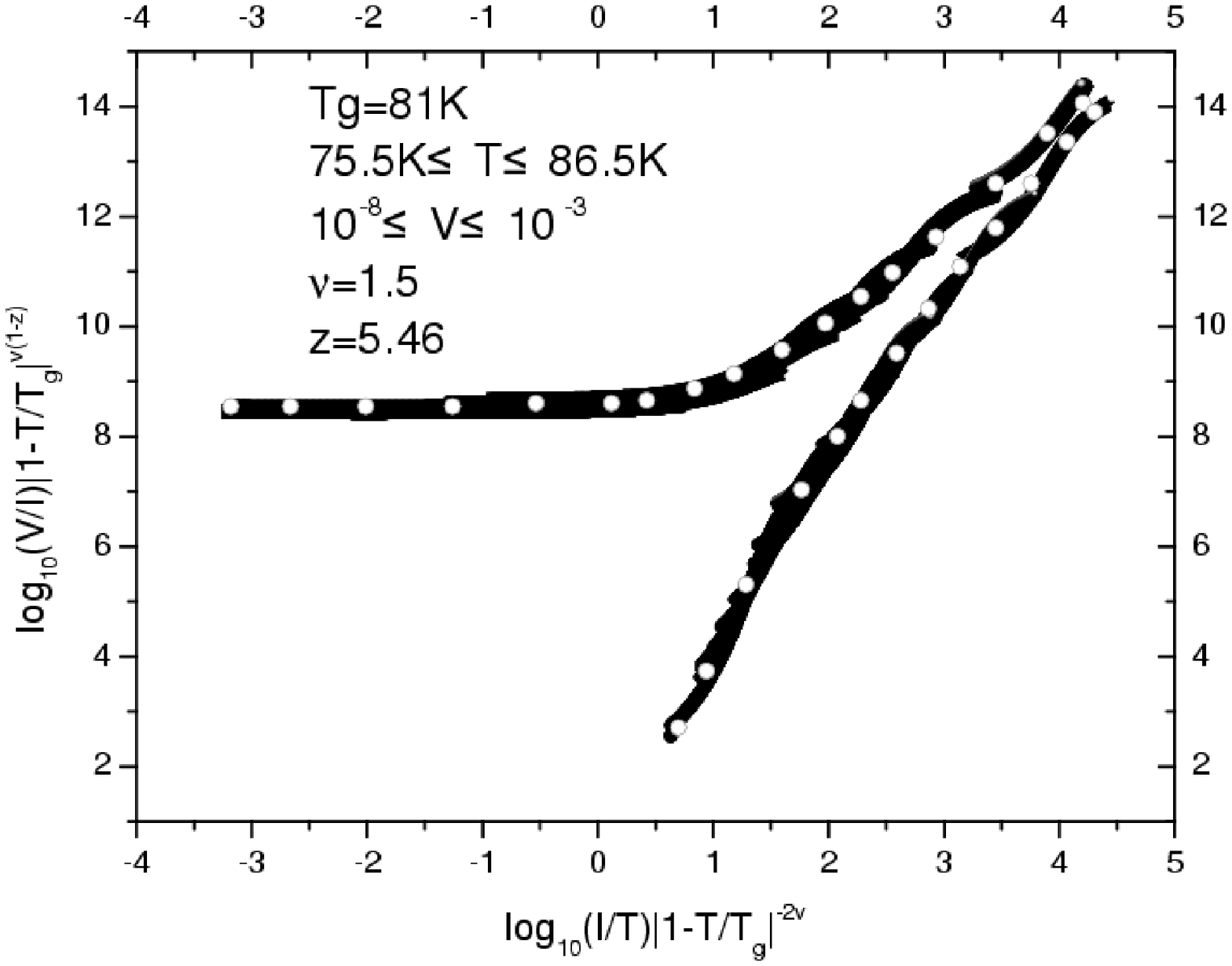}

\noindent (a)

\vspace{-3mm}

\includegraphics[width=0.585\linewidth,angle=270]{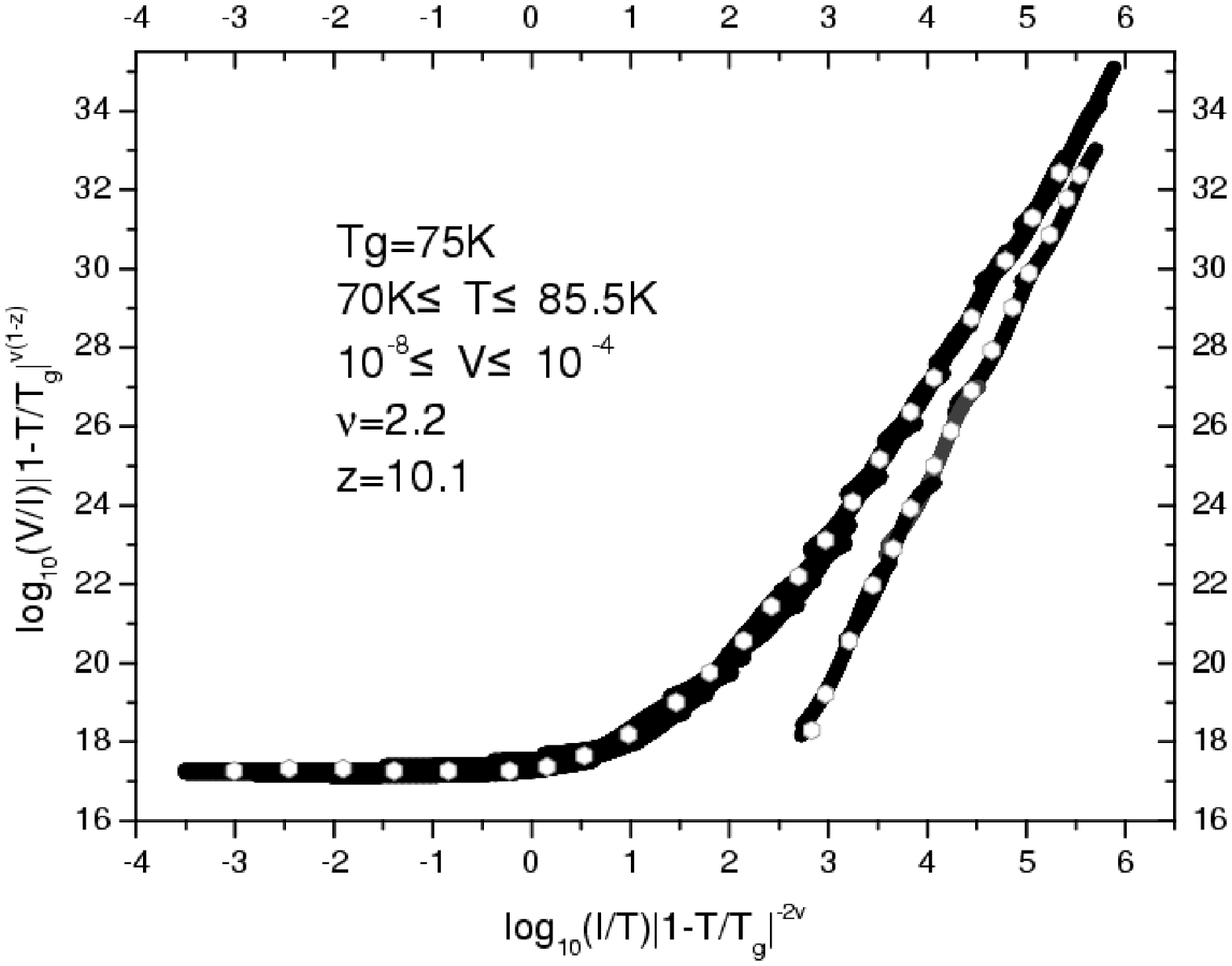}

\noindent (b)

\vspace{-3mm}

\includegraphics[width=0.55\linewidth,angle=270]{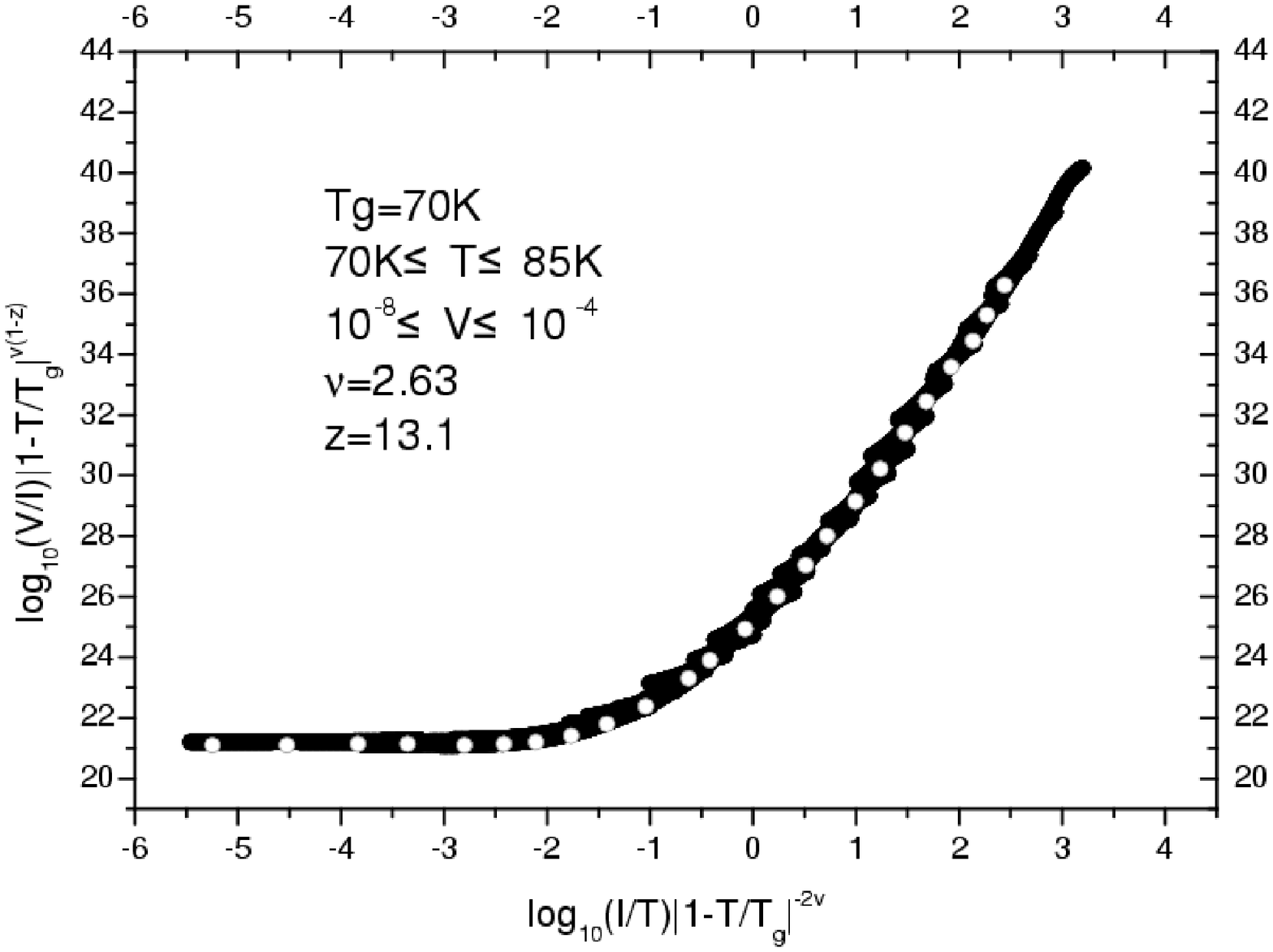}

\noindent (c)

\figcaption{$\bullet$: Collapse of the simulated I-V isotherms
reproduced from Eq.(\ref{eq16}) and Eq.(\ref{eq21})
using various critical parameters compared with\\
$\circ$: the experimental I-V isotherms for a 2200$\text{\AA}$
YBCO film in 4T measured by Strachan {\it et al.} \cite{Strachan}.
The conventional analysis is shown in (a).}
\end{figure}
On the basis of nearly 100 simulated $I-V$ isotherms from $68K$ to
$88K$ with temperature intervals of $0.2K$, we obtain three
different scaling data collapse analysis with assumed critical
temperature $T_{g}$ of $81K$,$75K$ and $70K$ respectively. In
Fig.2 we show the comparison of our result with the data collapses
of measured $I-V$ curves of Strachan et al. in Ref.
\cite{Strachan} and find fair agreement.

The analytical extended power law equations (\ref{eq16}),
(\ref{eq17}) and (\ref{eq19}) enable one to resolve the nontrivial
problem of determining the positive scaling function $\Gamma(v)$
between the average pinning force $\langle{\bf F}_p\rangle_t$ and
vortex velocity $v_L$ used in the VGFB model \cite{Vinokur} as
well as by Wang, Dong and Ting (WDT) \cite{WDT} in the form
$\langle{\bf F}_p\rangle_t=-\Gamma(v_L){\bf v_L}$. Thus, the
analytical equation for the nonlinear Hall resistivity can be
derived as
\begin{eqnarray}\nonumber
\rho_{xy}&=&\frac{\beta_0 \rho_{xx}^2}{\Phi_0
B}[\eta(1-\overline{C})-2\overline{C}\Gamma(v_L)]\\
&=&\mu_m \rho_n B \{(1+\overline{C})\frac{\rho_{xx}^2}{\rho_f^2}\
-2\overline{C}\frac{\rho_{xx}}{\rho_f}\}\text{,}\label{eq23}
\end{eqnarray}
with the longitudinal nonlinear resistivity defined in
Eqs.(\ref{eq26}) and (\ref{eq19}), $\rho_{xx}\simeq\rho_f
e^{-U_c/kT}$. The parameter $\overline{C}$ is defined as
$\overline{C}\equiv C(1-\overline{H}/H_{c2})$ with $\overline{H}$
the average magnetic field over the vortex core and $C$ is
describing the contact force on the surface of vortex core
\cite{WDT}. $\beta_0=\mu_m H_{c2}$ with $\mu_m=\tau e/m$ the
mobility of the charge carrier and $H_{c2}=\Phi_{0}/2\pi\xi^2$ the
upper critical magnetic field .

This equation is based on the conventional theories of Bardeen and
Stephen \cite{Bardeen} as well as Nozi\'{e}res and Vinen
\cite{Nozieres} for pinning free motion of vortex which predict
that the Hall effect stems from quasinormal core and hence has the
same sign as in the normal state. However, the microscopic
theories of the vortex dynamics in clean superconductors
\cite{KopninLopatin} predict that even in the absence of pinning,
the vortex contribution to Hall current may have the sign
different from the sign of Hall effect in the normal state. In
latter case, increasing pinning can remove the Hall anomaly and
even result in a second sign reversal of $\rho_{xy}$ observed in
some strongly anisotropic materials \cite{Kopnin, Ri, HagenLobb}.

With consideration of this factor, Hall resistivity equation
(\ref{eq23}) should be modified to the form
\begin{equation}\label{eq28}
\rho_{xy}(T)=\mu_m \rho_n B
\cdot\frac{A_s(T)}{A_n(T)}\{(1+\overline{C})\frac{\rho_{xx}^2}{\rho_f^2}\
-2\overline{C}\frac{\rho_{xx}}{\rho_f}\}\text{,}
\end{equation}
where $A_s(T)$ and $A_n(T)$ are the Hall force coefficients of the
sample in pinning free superconducting and normal states
respectively \cite{KopninLopatin}.

Besides the well-known puzzling scaling behavior
$\rho_{xy}\approx\rho_{xx}^\beta$, recently we find a new
universal scaling relation of the form
\begin{equation}
log_{10}\left|\frac{\rho_{xy}}{\rho_m}\right|\approx
\Phi\left(\frac{T-T_0}{T_m-T_0}\right)\text{,}
\end{equation}
where $\rho_m$ is the first extreme value of Hall resistivity from
low temperature side and $T_m$ is the temperature corresponding to
$\rho_m$. $T_0$ corresponds to the temperature where the Hall
resistivity become first measurable at the low temperature side.

\begin{figure}
\includegraphics[width=0.8\linewidth]{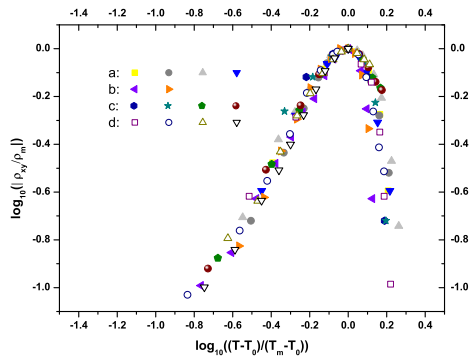}
\caption{Scaling of the anomalous $\rho_{xy}\sim T$ curves of
different kinds of cuprate superconductors in the form of
$log_{10}|\rho_{xy}/\rho_{m}|\sim
log_{10}((T-T_{0})/(T_{m}-T_{0}))$. Therein (a) represents
YBaCu$_3$O$_7$ \cite{Hagen, Xu}; (b) represents
Bi$_2$Si$_2$CaCu$_2$O$_{8+x}$ \cite{Ri}; (c) represents
HgBa$_2$CaCu$_2$O$_{6+\delta}$ \cite{Kang}; (d) represents
Tl$_2$Ba$_2$CaCu$_2$O$_8$ \cite{HagenLobb}.}
\end{figure}

This scaling relationship is consistent with the general Hall
resistivity equation(\ref{eq28}). In the case of YBCO (Fig.3(a)
\cite{Hagen, Xu}), the factor $A_s(T)/A_n(T)>0$, we see the Hall
anomaly at temperatures around $T_m$, on the other hand, for the
strongly anisotropic materials HBCCO(Fig.3(c) \cite{Kang}) and
TBCCO(Fig.3(d) \cite{HagenLobb}), we see the double sign reversal
of $\rho_{xy}(T)$, perhaps owing to that $A_s(T)/A_n(T)<0$, as
mentioned by Kopnin and Vinokur \cite{Kopnin}.

In conclusion, we demonstrate that the longitudinal $E(J)$
response have the general form of extended power law. This
nonlinear behavior naturally explains the scaling of longitudinal
and Hall resistivities near the vortex glass transition.

This work is supported by the Ministry of Science \& Technology of
China (NKBRSG-G 1999064602) and the National Natural Science
Foundation of China under Grant No.10174003, No.50377040 and
No.90303008.

\end{document}